\documentclass{mem}
\usepackage{natbib}\usepackage{txfonts}\usepackage{balance}
\usepackage{graphicx}
\usepackage[a4paper,breaklinks,dvipdfm]{hyperref}
\idline{75}{282}
\begin{document}
\def\teff{$T\rm_{eff }$}
\def\kms{$\mathrm {km s}^{-1}$}
\newcommand{\omc}{\mbox{$\omega$ Cen~}} 
\newcommand{\msun}{$M_{\odot}\,$}

\title{The soundtrack of RR Lyrae in $\omega$ Cen at high-frequency\thanks{Based on data collected 
with  ULTRACAM@NTT (La Silla, ESO, PID: 087.D-0216)}}

   \subtitle{}

\author{
A. \,Calamida\inst{1} 
\and S. K. Randall\inst{2}
\and M. Monelli\inst{3}
\and G. Bono\inst{4} 
\and R. Buonanno\inst{4}
\and G. Strampelli\inst{4}
\and M. Catelan\inst{5}
\and V. Van Grootel\inst{6}
\and M. L. Alonso\inst{5}
\and P. B. Stetson\inst{7}
\and R. F. Stellingwerf\inst{8}}

 
\institute{
Space Telescope Science Institute - AURA
3700 San Martin Drive
Baltimore, MD 21218,
USA \email{calamida@stsci.edu}
\and European Southern Observatory, Karl-Schwarzschild-Str. 2, 85748 Garching bei Muenchen, Germany
\and Instituto de Astrofisica de Canarias, Calle Via Lactea, E38200 La Laguna, Tenerife, Spain
\and Universit\`a di Roma Tor Vergata, Via della Ricerca Scientifica 1, 00133 Rome, Italy 
\and Pontificia Universidad Cat\'olica de Chile, Av. Vicu\~{n}a Mackenna 4860, 782-0436 Macul - The Milky Way Millennium Nucleus, Santiago, Chile
\and Institut d'Astrophysique et de G\' eophysique, Universit\' e de Li\` ege, 17 All\'ee du 6 Ao\^ ut, B-4000 Li\` ege, Belgium
\and NRC-HIA, 5071 West Saanich Rd, Victoria, BC V9E 2E7, Canada
\and Stellingwerf Consulting, 11033 Mathis Mountain Rd SE, Huntsville, AL, USA}

\authorrunning{Calamida}

\titlerunning{The soundtrack of $\omega$ Cen RR Lyrae stars at high-frequency}

\abstract{We present preliminary Sloan $u',g'$-band light curves for a sample of known RR Lyrae variables in the Galactic globular cluster 
$\omega$ Cen. Results are based on the partial reduction of multi-band time series photometric data collected during
six consecutive nights with the visitor instrument ULTRACAM mounted on the New Technology Telescope (La Silla, ESO). 
This facility allowed us to simultaneously observe in three different bands (Sloan $u',g',r'$) a field of view 
of $\sim$ 6$\times$6 arcminutes. The telescope and the good seeing conditions 
allowed us to sample the light curves every 15 seconds. We ended up with a data set of $\sim$ 6,000 
images per night per filter, for a total of more than 200,000 images of the selected field.
This data set allowed us to detect different kind of variables, such as RR-Lyraes, SX Phoenicis, 
eclipsing binaries, semi-regulars. More importantly, we were able for the first time to sample at 
high-frequency cluster RR Lyraes in the $u',g'$-band and to show in detail the pulsation phases 
across the dip located along the rising branch of RR-Lyraes.
}

\maketitle{}

\section{Introduction and data reduction}
The Galactic Globular Cluster (GGC) \omc is 
the most massive known in our Galaxy (2.5 $\times 10^6$ \msun, van de Ven et al. 2006) and the
only one known to host at least three separate stellar populations with a large 
undisputed spread in metallicity (Norris \& Da Costa 1995; Smith et al.\ 2000, Calamida et al.\ 2009, Johnson et al.\ 2010).
Moreover, empirical evidence suggests the occurrence of a He-enhanced
stellar population in this cluster (Bedin et al.\ 2004).

Even though $\omega$ Cen presents properties which still need to be properly understood, its stellar content 
is a gold mine to investigate several open problems concerning 
stellar evolution and its dependence on the metallicity. This may apply not only 
to red giant, main sequence and horizontal branch stars, but also to RR Lyrae stars.
The near-infrared period-luminosity relation for RR Lyrae stars is one of the most important methods of
estimating distances and calibrating distance indicators (Bono et al.\ 2001; Cassisi et al.\ 2004).
Unfortunately, the number of Galactic calibrator with accurate trigonometric parallaxes for RR Lyrae
stars is restricted to a single object, RR Lyr itself. The best way to overcome this problem and accurately 
study the properties of these robust distance indicators is to observe them in GGC, as the 
RR Lyrae stars in these systems are located at about the same distance and are affected by the same reddening. 
In particular, \omc is the perfect target since it hosts more than 186 RR Lyraes (86 RRab, 100 RRc, Kaluzny et al.\ 2004, hereafter KA04)
with a metal content ranging from $[Fe/H] \sim$ -2 to -1 (Rey et al.\ 2000, using 131 objects; Sollima et al.\ 2006, using 74 objects).
Moreover, the \omc distance has been estimated using different distance indicators such as 
the eclipsing binary OGLE 17 (Thompson et al.\ 2001), the $V$-band metallicity relation of RR Lyrae stars (Rey et al.\ 2000; Catelan et al. 2004), 
and a dynamical analysis by van de Ven et al. (2006), based on proper motions and radial velocities of a large sample of individual stars.
\begin{figure}[]
\resizebox{\hsize}{!}{\includegraphics[clip=true]{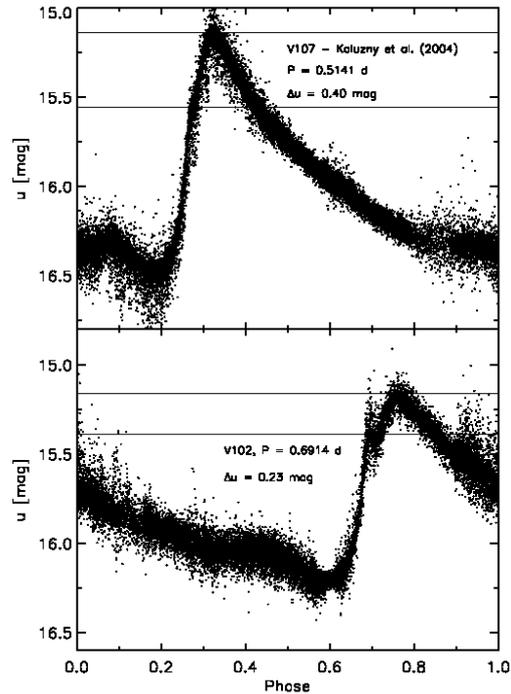}}
\caption{\footnotesize
Phased $u$-band light curves for two fundamental mode RR Lyrae stars, $V107$ (top panel) and
V102 (bottom).
The solid lines mark the location of the maximum and of the dip along the light curves.}
\label{fig1}
\end{figure}
We present here preliminary results based on the partial reduction of Sloan $u', g', r'$-band (hereafter $u,g,r$-band) 
time series data collected during six consecutive nights (April 22-27, 2011) with the visitor instrument ULTRACAM mounted 
on the New Technology Telescope (NTT, La Silla, ESO). ULTRACAM is a CCD camera designed to provide imaging photometry 
at high temporal resolution in three different bands simultaneously.
The exposure time for each image is $15$ s and about $6,000 \times 2$ (chip1 + chip2) images per filter per night have been collected, 
for a total of $\approx$ $216,000$ images. The pointing was chosen in order to perform asteroseismology of the newly discovered 
O-type variable sub-dwarfs (sdO) in \omc (Randall et al.\ 2011) and it is a field of view of $\sim$
6'$\times$6' South-East of the cluster center. The average seeing during the observations ranged from 1.0 to 1.7 arcsec 
(data for April 26 were discarded, due to the poor atmospheric conditions of the night).
The Point-Spread-Function (PSF) photometry on all images has been performed with DAOPHOTIV /ALLSTAR (Stetson 1987), 
and ALLFRAME (Stetson 1994) has been adopted to perform PSF photometry on the images of each chip, each band and 
each night simultaneously. The catalogs of each night have been scaled to the assumed reference night, April 25, and matched 
with our astrometrized Wide Field Imager (WFI) Ð- Advanced Camera for Surveys (ACS) catalog, which provided us with 
$F435W, F625W, F658N, U,B,V,I$-band photometry for each star (Castellani et al.\ 2007).
The reduction of the $u$-band images is complete, while we partially reduced the $g$-band images (April 25) and did not
start the reduction of $r$-band images yet.

\section{Light curves}
Variables have been selected by adopting the Welch-Stetson index (Welch \& Stetson 1993) and $909$ candidates were identified 
in chip1 (the closest to the cluster center) and $562$ on chip2. Several candidates were then selected by eye, $21$ on chip1 and $12$
 on chip2. In the ULTRACAM field of view fall 23 known variables by KA04.
Our selection includes 5 known RRab, 7 RRc, 1 SX Phoenicis, 2 eclipsing binaries, 1 field variable, and 17 new 
candidate variables, including a new hot horizontal branch pulsator (these results will be presented in detail in a forthcoming paper).

Fig.~1 shows the $u$-band phased light curves of two RRab, $V107$ (top panel) and $V102$ (bottom) 
based on five nights of observations. Note that magnitudes are not calibrated.
The periods have been estimated by using the NASA Exoplanet Archive 
Periodogram service and by adopting a Lomb-Scargle algorithm\footnote{Visit the following URL for more details: http://exoplanetarchive.ipac.caltech.edu/cgi-bin/Periodogram/nph-simpleupload}. Thanks to the high temporal resolution of ULTRACAM data, we were able to fully sample the rising branch of the variables, 
with sufficient accuracy (the photometric error on the single measurement being $\sigma_u \approx$ 0.02 mag) to disclose interesting details that could not be 
previously observed in the KA04 light curves. 
The dip along the rising branch of the $u$-band light curves is clearly 
visible, as observed by Lub (1977) and  modeled with hydrodynamical pulsation models by Bono \& Stellingwerf (1994) for the 
high-amplitude RR Lyrae stars.
The dip is more pronounced for $V102$, which has a longer period, P = 0.691 d,  compared to $V107$ for which P = 0.514 d. 
The difference between the phase of the maximum and the dip attain very similar values, within current errors, 
for all the four RRab, i.e. $\Delta \phi \approx 0.04 - 0.06$. 
On the other hand, the magnitude difference between the maximum and the dip of the light curves  
is decreasing for increasing pulsation period.  The magnitude difference for variables V107 and
V102 is  $\sim$ 0.40 and $\sim$ 0.23 mag, with periods of 0.5141 d and 0.6914 d, respectively, 
while for the other two RRab is 0.28 ($V41$, P = 0.693 d\footnote{For variables $V41$ and $V350$ 
periods estimated by KA04 have been adopted.}) and 0.50 mag  ($V113$, P = 0.572 d). 
$V112$ is missing because we were not able to observe a complete cycle for 
this star. There is then a preliminary evidence that the magnitude difference between
the maximum and the dip correlates with the period. 
A more complete analysis will be presented in a forthcoming paper.
It is worth mentioning that the dip is also present but barely visible in the $g$-band light curves of the same RRab stars.

Fig.~2 shows the $u$- (top panel) and $g$-band (bottom) light curve for RRc $V147$. 
This is a mixed-mode variable according to KA04. The high temporal sampling together with the photometric 
accuracy of the single measurements, $\sigma_g \approx$ 0.015 mag, allowed us to clearly identify a dip
in the $g$-band light curve.
The difference in magnitude between the maximum and the dip is  $\sim$ 0.15 mag and the phase difference is $\sim$ 0.2.
The same dip is identified also on the $u$-band light curve.
The difference in magnitude between the maximum and the dip is $\sim$ 0.10 mag, with the same phase difference of $\sim$ 0.2.

\begin{figure}[]
\resizebox{\hsize}{!}{\includegraphics[clip=true]{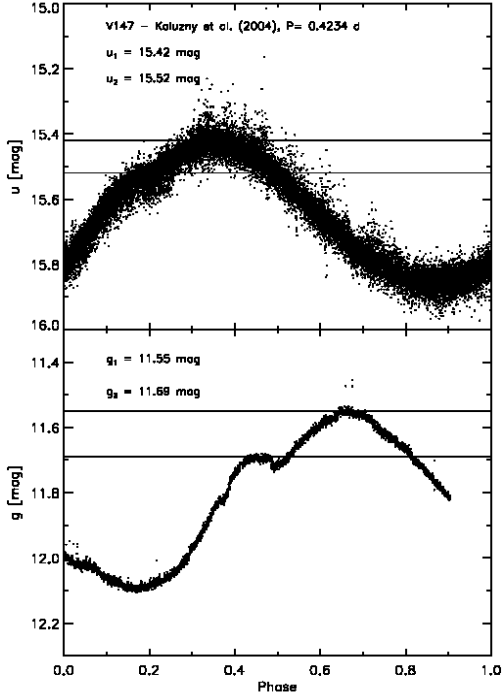}}
\caption{\footnotesize $u$- (top panel) and $g$-band (bottom) 
light curve for RRc $V147$. Solid lines mark the location of the maximum and of 
the dip along the rising branch of the  light curves.}
\label{fig2}
\end{figure}

Fig.~3 shows the Bailey diagram in both bands for our RR Lyrae stars. RRab $V41$  and $V113$ are 
missing in the $g$-band since we were not able to estimate the amplitude with only one night of observations.
The stars split in two well defined groups, one including only fundamental mode RR Lyraes ($A_u \approx §A_g > $ 0.7) and the other one including
the first overtone RR Lyraes ($A_u \approx A_g < $ 0.7). The paucity of the sample does not allow us to establish the presence of any clear trend.
On the other hand, it is worth noting that all the RRc pulsators showing a clear dip along the rising branch of 
both the $u$- and the $g$-band light curves  ($V110, V145, V147$) and the candidate mixed-mode 
RRc pulsator $V350$ (Olech et al.\ 2009) have larger amplitudes, 
ranging from $A_u \sim$0.4 to $\sim$ 0.6 and $A_g \sim$0.33 to $\sim$ 0.43 mag. 

\begin{figure}[]
\resizebox{\hsize}{!}{\includegraphics[clip=true,angle=90]{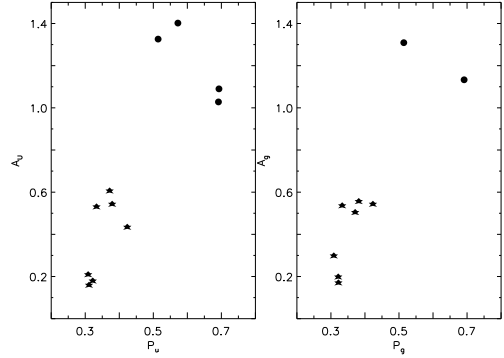}}
\caption{\footnotesize
Bailey diagram for 11 out of the 12 RR Lyrae stars in the $u$-band (left) and
9 out 12 stars for the $g$-band (right). 
}
\label{fig4}
\end{figure}

\section{Conclusions}
We presented preliminary Sloan $u,g$-band light curves of known fundamental and first overtone 
RR Lyrae variables in $\omega$ Cen.
The high temporal sampling of ULTRACAM data allowed us for the first time to clearly
identify in the $u$- and $g$-band light curves the dip located along the rising branch of RR Lyraes 
and a few mixed-mode pulsators.

After concluding the data reduction for the $g$- and $r$-band time-series, 
we plan to analyze more in detail the sample of RR Lyraes to look at the occurrence of the dip in different
mode pulsators and its correlations with periods and amplitudes in different photometric bands.
Furthermore, we will analyze the candidate mixed-mode pulsators to better constrain their 
pulsational properties.

\begin{acknowledgements}
We kindly acknowledge the ULTRACAM team, in particular Thomas Marsh and Vik Dhillon.
\end{acknowledgements}

\bibliographystyle{aa}

\end{document}